\documentstyle[12pt]{article}

\begin{document}
\hspace*{1.5cm} Running head: Stable periodic waves in coupled KS-KdV equations\\
\break
\break
\begin{center}
{\bf Stable Periodic Waves in Coupled Kuramoto-Sivashinsky~--~\\
Korteweg-de Vries Equations\\}
\vspace*{1cm}
Bao-Feng FENG, Boris A. MALOMED${}^{1 \ast}$
 and Takuji KAWAHARA${}^{2 \ast\ast}$ \\
\vspace*{0.5cm}
{\it Department of Mathematics, the University of Kansas,\\
Lawrence, KS 66045,  U.S.A. \\

${}^1$Instituto de Fisica Teorica~--~UNESP, R. Pamplona 145,\\ 01405-900 S\~{a}o Paulo, Brazil\\

${}^2$Department of Aeronautics and Astronautics, Graduate School of \\Engineering, Kyoto University, Sakyo-ku, Kyoto 606-8501\\}
\end{center}

\hspace*{5cm}(Received $\qquad\quad$ 2002)\\
\vspace*{1.25cm}
\par
Periodic waves are investigated in a system composed of a
Kuramoto~-~Sivashinsky -- Korteweg~-~de Vries (KS-KdV) equation linearly
coupled to an extra linear dissipative one. The model describes, e.g., a
two-layer liquid film flowing down an inclined plane. It has been recently
shown that the system supports stable solitary pulses. We demonstrate that a
perturbation analysis, based on the balance equation for the net field
momentum, predicts the existence of stable cnoidal waves (CnWs) in the same
system. It is found that the mean value $u_{0}$ of the wave field $u$ in the
main subsystem, but {\em not} the mean value of the extra field, affects the
stability of the periodic waves. Three different areas can be distinguished
inside the stability region in the parameter plane ($L$,$u_{0}$), where $L$
is the wave's period. In these areas, stable are, respectively, CnWs with
positive velocity, constant solutions, and CnWs with negative velocity.
Multistability, i.e., the coexistence of several attractors, including the
waves with several maxima per period, appears at large value of $L$. The
analytical predictions are completely confirmed by direct simulations.
Stable waves are also found numerically in the limit of vanishing
dispersion, when the KS-KdV equation goes over into the KS one.
\break
\break
${}^{\ast}$ Present address: Department of Interdisciplinary Studies, Faculty of Engineering, Tel Aviv University, Tel Aviv 69978, Israel \\
${}^{\ast\ast}$ E-mail: kawahara@impact.kuaero.kyoto-u.ac.jp\\
\newpage
\baselineskip=16pt
\hspace*{-0.7cm}{\bf \S 1.  Introduction}\\

It is commonly known that periodic waves, as well as their solitary-wave
(SW) counterparts, are coherent structures of fundamental importance in the
study of physical systems governed by nonlinear evolution equations \cite
{Hohen93}. Stability of these solutions under small perturbations is an
issue of basic interest in its own right and for applications, that is why
it has been investigated in many works (see, e.g., refs. \cite{Takuji3}--\cite{JPSJ} and references therein). In particular, the
stability of periodic waves in perturbed Korteweg - de Vries (KdV) and
Benjamin-Ono equations was investigated, respectively, in refs. \cite
{Spector86}--\cite{Bar} and \cite{KaupPF,Spector4}.

An important one-dimensional wave-generating model that combines
conservative and dissipative effects is a mixed Kuramoto~-~Sivashinsky (KS)
-- KdV equation, 
\begin{equation}
u_{t}+uu_{x}+u_{xxx}=-\alpha u_{xx}-\gamma u_{xxxx}\,,  \label{Benney}
\end{equation}
with $\alpha ,\gamma >0$, which was first introduced by Benney \cite{Benney}
and is often called the Benney equation. This equation finds various
applications in plasma physics, hydrodynamics and other fields \cite{KSKdV}--\cite{CV95}.
Localized structures are, obviously, important objects in systems of this
type \cite{SP}--\cite{CDK95}; however, SWs cannot be stable in the Benney equation proper,
as the zero solution, which is a background on top of which SWs are to be
found, is linearly unstable in this equation due to the presence of the
linear gain, which is accounted for by the coefficient $\alpha $ in eq. (\ref
{Benney}) \cite{JPSJ}. A stabilized version of the Benney equation was
recently proposed by the present authors in ref. \cite{Malomed01}. It is
based on the KS-KdV equation for a wave field $u(x,t)$, which is linearly
coupled to an additional linear dissipative equation for an extra field 
$v(x,t)$, that provides for the stabilization of the zero background: 
\begin{eqnarray}
u_{t}+uu_{x}+u_{xxx}-v_{x} &=&-\alpha u_{xx}-\gamma u_{xxxx},  \label{u} \\
v_{t}+cv_{x}-u_{x} &=&\Gamma v_{xx}.  \label{v}
\end{eqnarray}
Here, $\alpha $, $\gamma $ and $\Gamma $ are positive coefficients
accounting for the gain and loss in the $u$-subsystem and loss in the 
$v$~-~subsystem, respectively, and $c$ is a group-velocity 
mismatch between the
fields. The system describes, for instance, the propagation of surface waves
in a two-layer liquid film in the case when one layer is dominated by
viscosity (see a review \cite{RMP} and ref. \cite{Malomed01} and references
therein).

In the work \cite{Malomed01}, it was shown that the stability of the zero
solution to eqs. (\ref{u}) and (\ref{v}) is compatible with the existence of
stable SWs. Actually, SW solutions for this system were obtained, with a
high accuracy, in an approximate analytical form by means of a perturbation
theory based on a balance equation for the net wave momentum. The existence
and stability of these solitary pulses was confirmed by direct simulations.
In fact, it is the first example of stable SWs in models of the KS type.

In contrast to SWs, periodic waves can be stable in the Benney equation \cite
{Takuji3}--\cite{KT89},\cite{Kawahara83}--\cite{KSpulse2}. Periodic waves are more likely than
SWs to be stable in dissipative nonlinear systems because the stability of
the zero solution is not a necessary condition in this case. Furthermore, in
experiments it is easier to observe periodic waves than SWs, this is why
detailed study of their existence and stability is a relevant problem.

A purpose of this work is to find periodic-wave solutions to the system of
eqs. (\ref{u}) and (\ref{v}), and to investigate their stability. In \S
2, we find a family of exact cnoidal-wave (CnW) solutions to a zero-order
system, in which the gain and loss are dropped. In \S 3, a perturbation
theory for the periodic waves is developed by treating the gain and loss as
small perturbations. To this end, the balance equation for the net field
momentum is employed, following the technique developed for SWs in ref. \cite
{Malomed01}. We demonstrate that a condition for equilibrium between the
gain and loss may select two (or none) particular steady-state CnW solutions
out of the continuous family existing in the zero-order system. In the case
when two solutions are selected, the one with a larger amplitude is stable,
while the other one is unstable, playing the role of a separatrix bordering
an attraction domain of the stable CnW.

In \S 4, we present results of direct numerical simulations of the full
system, which turn out to be in very good agreement with the analytical
predictions. In \S 5, we briefly consider the special case of the
system with the vanishing dispersion, i.e., with the KS-KdV equation (\ref{u})
replaced by the KS one. In this case, direct simulations demonstrate the
existence of stable periodic waves. The paper is concluded by \S 6.\\

\hspace*{-0.7cm}{\bf \S 2.  Cnoidal Waves in the Zero-Order System}\\

In this work, we focus on the study of steady traveling-wave solutions to
eqs. (\ref{u}) and (\ref{v}) in a spatially periodic domain $0<x<\,L$ with
boundary conditions (b.c.) 
\begin{eqnarray}
u(0) &=&u(L),\quad u_{x}(0)=u_{x}(L),\quad u_{xx}(0)=u_{xx}(L),\quad
u_{xxx}(0)=u_{xxx}(L);  \nonumber \\
v(0) &=&v(L),\quad v_{x}(0)=v_{x}(L).\qquad \qquad \qquad \qquad \qquad
\qquad  \label{bcs}
\end{eqnarray}
Due to the imposed periodic b.c., the stability results obtained below (in
\S 4) by means of direct simulations are limited to perturbations which
have the same periodicity as the unperturbed solution. In a long system, a
periodic wave may be subject to extra instabilities against
periodicity-breaking perturbations. However, in this work we will display
generic examples of stable waves with a fundamental period $L/N$ rather than 
$L$, with $N$ taking values, typically, $3$ or $4$. Those examples will
demonstrate that, as a matter of fact, the waves may be stable against
periodicity-breaking perturbations too.

The group-velocity parameter $c$ may be set equal to zero in eq. (\ref{v}),
as a substitution 
\begin{equation}
u\rightarrow u+u_{0},\quad x\rightarrow x+u_{0}t  \label{trf}
\end{equation}
transforms eqs. (\ref{u}) and (\ref{v}) into the same system, with $c$
replaced by $c-u_{0}$. In other words, we can eliminate $c$ by subtracting
the constant (background) value $u_{0}\equiv c$ from the field $u(x,t)$
(obviously, this transformation could not be applied in the case of SWs, for
which the zero value of the background must be kept). Therefore, $c$ is
assumed to be zero hereafter.

In the zero-order approximation, corresponding to 
$\alpha =\gamma =\Gamma =0$,
eqs. (\ref{u}) and (\ref{v}) take the form of a conservative system
consisting of the KdV equation coupled to an extra linear one: 
\begin{equation}
u_{t}+uu_{x}+u_{xxx}=v_{x},{\quad }v_{t}=u_{x}  \label{zero-order}
\end{equation}
(recall we have set $c=0$). In ref. \cite{Malomed01} it was demonstrated
that the conservative system (\ref{zero-order}) is {\em not} an integrable
one.

It is well known that the KdV equation proper, 
\begin{equation}
u_{t}+uu_{x}+u_{xxx}=0,  \label{KdV}
\end{equation}
admits a family of periodic-wave solutions that have an arbitrary period $L$
and travel at an arbitrary velocity $s$. These solutions are often called
cnoidal waves (CnWs), which stems from the symbol ${\rm cn}$ for the
elliptic cosine. Making use of definitions 
\begin{equation}
\xi \equiv x-st,\,\,\eta L\equiv 2K(m),  \label{KdV-etaL}
\end{equation}
CnW solutions to the KdV equation can be cast in the form 
\begin{equation}
u(x,t)=12\eta ^{2}\left[ {\rm dn}^{2}(\eta \xi ,m)-E(m)/K(m)\right] ,
\label{KdV-cnoi}
\end{equation}
\begin{equation}
s=4\eta ^{2}\left[ (2-m)-3E(m)/K(m)\right] .  \label{KdV-cnois}
\end{equation}
Here, $\eta $ is an intrinsic parameter of the solution family, which
determines the amplitude and shape of the wave, ${\rm dn}(\xi ,m)$ is the
Jacobi's elliptic function with the modulus $q\equiv \sqrt{m}$, and $K(m)$
and $E(m)$ are complete elliptic integrals of the first and second kinds. In
the limit of $m\rightarrow 1$, the CnW solution goes over into the KdV
soliton. Note that the solution (\ref{KdV-cnoi}) has a zero mean value of 
$u(\xi )$.

On the basis of the CnW solutions to the KdV equation presented above, it is
easy to generate a family of periodic-wave solutions to the zero-order
system (\ref{zero-order}), making use of the same definitions (\ref{KdV-etaL})
as above. The result is 
\begin{equation}
u(x,t)=u_{0}+12\eta ^{2}\left[ {\rm dn}^{2}(\eta \xi ,m)-E(m)/K(m)\right] ,
\label{cnoi-u}
\end{equation}

\begin{equation}
v(x,t)=v_{0}-12\frac{\eta ^{2}}{s}\left[ {\rm dn}^{2}(\eta \xi ,m)-E(m)/K(m)
\right] ,  \label{cnoi-v}
\end{equation}
where the wave's velocity $s$ is related to the mean value $u_{0}$ of the 
$u$~-~field as follows: 
\begin{equation}
u_{0}+4\eta ^{2}\left[ (2-m)-3E(m)/K(m)\right] =s-\frac{1}{s}\,.
\label{etas}
\end{equation}
Lastly, $v_{0}$ is the mean value of the $v$-field.\\

\hspace*{-0.7cm}{\bf \S 3.  The Perturbation Theory for Cnoidal Waves}\\

In this section, we perform a perturbation analysis of the full system of
eqs. (\ref{u}) and (\ref{v}) in the presence of the gain and loss terms.
First, we notice that eqs. (\ref{u}) and (\ref{v}) conserve two ``masses'', 
\[
M=\int_{0}^{L}u(x)\,dx,\quad N=\int_{0}^{L}v(x)\,dx 
\]
(which indeed have the meaning of masses in the application to liquid-film
flows, see refs. \cite{RMP,Malomed01} and references therein). The net field
momentum $P$ for periodic solutions is defined as 
\begin{equation}
P=\frac{1}{2}\int_{0}^{L}\left[ u^{2}(x)+v^{2}(x)\right] dx.
\label{momentum}
\end{equation}
Unlike the masses, the momentum is not conserved by eqs. (\ref{u}) and (\ref
{v}), but it is conserved in the zero-order approximation, setting $\alpha
=\gamma =\Gamma =0$, therefore the balance equation for $P$ may be naturally
used as a basis for the simplest perturbation theory \cite{Malomed01}.

With regard to the periodic boundary conditions (\ref{bcs}), an {\em exact}
evolution (balance) equation for $P$ can be easily derived: 
\begin{equation}
\frac{dP}{dt}=\int_{0}^{L}\left( \alpha u_{x}^{2}-\gamma u_{xx}^{2}-\Gamma
v_{x}^{2}\right) \,dx.  \label{dPdtp}
\end{equation}
In the first-order approximation of the perturbation theory, we assume that
the system inherits the unperturbed CnW solutions in the form of eqs. (\ref
{cnoi-u}) and (\ref{cnoi-v}), while the amplitude parameter $\eta $ may vary
as a function of the slow time, $T\equiv \epsilon t$. In particular, the
momentum corresponding to the unperturbed solution is 
\begin{equation}
P_{0}=\frac{1}{2}\left( u{_{0}}^{2}+v_{0}^{2}\right) L+48\eta ^{3}\left( 1+
\frac{1}{s^{2}}\right) .  \label{P0}
\end{equation}
Calculating the integral on the r.h.s. (right-hand side) of eq. (\ref{dPdtp})
analytically and substituting the expression (\ref{P0}) in l.h.s., we
derive an evolution equation for the amplitude, 
\begin{equation}
\frac{d\eta }{dT}=-\frac{6\eta ^{3}s^{2}\left( s^{2}+1\right) }{105c_{0}
\left[ 3\left( s^{2}+1\right) ^{2}-4\left( s^{2}-u_{0}s-1\right) \right]}
\left[ 80\gamma c_{2}\eta ^{2}+28c_{1}\left( \alpha -\Gamma /s^{2}\right)
\right] ,  \label{dPdt}
\end{equation}
where we define 
\begin{equation}
\begin{array}{ccc}
c_{0} & \equiv & (m-1)(3m-2)K(m)-2(1-2m)E(m) \\ 
&  & -3\left[ E(m)+(m-1)K(m)\right] ^{2}/K(m), \\ 
c_{1} & \equiv & (m-1)(m-2)K(m)-2(1-m+m^{2})E(m), \\ 
c_{2} & \equiv & (m-1)(m^{2}+2m-2)K(m)-(2-3m-3m^{2}+2m^{3})E(m).
\end{array}
\label{C}
\end{equation}

In the subsequent analysis, it is more convenient to follow the evolution of
the wave's velocity $s$, rather than the amplitude $\eta $. To this end, we
express $\eta $ in terms if $s$, making use of eq. (\ref{etas}): 
\begin{equation}
4\eta ^{2}=\frac{s-1/s-u_{0}}{(2-m)-3E(m)/K(m)}\,,  \label{etadel}
\end{equation}
a corollary of which is 
\begin{equation}
\frac{ds}{d\eta }=\frac{8\eta \left[ (2-m)-3E(m)/K(m)\right] }{1+1/s^{2}}\,.
\label{dsdeta}
\end{equation}
Using eqs. (\ref{etadel}) and (\ref{dsdeta}), we derive, after some algebra,
the following evolution equation for $s$ from eq. (\ref{dPdt}): 
\begin{equation}
\frac{ds}{dT}=\frac{Cs^2(s-1/s-u_{0})^{2}} {3\left(s^2+1\right)^2-4
\left(s^2-u_{0}s-1\right)} (s^{3}+a_{2}s^{2}-s-a_{0}),  \label{dsdt}
\end{equation}
where 
\begin{eqnarray}
C &\equiv &-\frac{4\gamma c_{2}}{7c_{0}[(2-m)-3E(m)/K(m)]^{2}},  \nonumber \\
&& 
\begin{array}{ccc}
a_{2} & \equiv & -u_{0}+1.4\left( {\alpha }/{\gamma }\right) c_{1}\left[
(2-m)-3E(m)/K(m)\right] /c_{2}, \\ 
a_{0} & \equiv & 1.4\left( {\Gamma }/{\gamma }\,\right) c_{1}\left[
(2-m)-3E(m)/K(m)\right] /c_{2},
\end{array}
\label{a}
\end{eqnarray}
and the constants $c_{0},c_{1},c_{2}$ were defined in eq. (\ref{C}). 

A steady state providing for equilibrium between the gain and loss is a
fixed point (FP) of eq. (\ref{dsdt}), which is attained when r.h.s. of the
equation vanishes. This is possible either if $s$ is a root of the cubic
equation 
\begin{equation}
s^{3}+a_{2}s^{2}-s-a_{0}=0,  \label{cubic}
\end{equation}
or if $s-1/s-u_{0}=0$, i.e., 
\begin{equation}
s=\frac{1}{2}\left( u_{0}\pm \sqrt{u_{0}^{2}+4}\right) \,.  \label{trivial}
\end{equation}
The FPs (\ref{trivial}) lead to a trivial result, as it immediately follows
from eq. (\ref{etas}) that they corresponds to a constant solution with 
$\eta =0$. Therefore, we now focus on eq. (\ref{cubic}). In fact, precisely
the same equation (\ref{cubic}) can also be obtained from the solvability
condition of a standard asymptotic perturbation analysis of the full system
(we do not present it here).

Physical roots of eq.  (\ref{cubic}) are those which are not only real but
also provide for $\eta ^{2}>0$ through eq. (\ref{etadel}). Noting that the
combination $(2-m)-3E(m)/K(m)$ in eq. (\ref{etadel}) 
is negative if $m<0.96$, and positive if $m>0.96$, 
we conclude that physical roots must belong to
the following intervals, in order to guarantee $\eta ^{2}>0$: 
\begin{equation}
\begin{array}{ccc}
s<\frac{1}{2}(u_{0}-\sqrt{u_{0}^{2}+4}),\,{\rm or}\,\,0<s<\frac{1}{2}(u_{0}+ 
\sqrt{u_{0}^{2}+4}), &  & {\rm if}\ \,m<0.96; \\ 
\frac{1}{2}(u_{0}-\sqrt{u_{0}^{2}+4})<s<0,\ {\rm or}\,\ s>\frac{1}{2}(u_{0}+ 
\sqrt{u_{0}^{2}+4}), &  & {\rm if}\ \ m>0.96.
\end{array}
\label{cond1}
\end{equation}
In particular, for $u_{0}=0$, physical roots lie in the intervals $s<-1$ or 
$0<s<1$ if $m<0.96$, and $-1<s<0$ or $s>1$ if $m>0.96$.

Further, we define $q=-1/3-\left( a_{2}/3\right) ^{2}$ and $\,r=\left(
1/6\right) (3a_{0}-a_{2})$ for eq. (\ref{cubic}). Then, as is well known,
roots of the cubic equation can be classified as follows: there are
one real root and a pair of complex conjugate ones if 
$q^{3}+r^{2}>0$, and three real roots if $q^{3}+r^{2}\leq 0$.
Thus, generally speaking, there might exist up to three physical roots to
eq. (\ref{cubic}). However, a detailed numerical analysis based on the above
formulas has only revealed cases with two physical roots, or none. In
particular, if there is a single real root of the cubic equation 
(\ref{cubic}), this root is always unphysical, as it does not satisfy the
above-mentioned condition $\eta ^{2}>0$. Two other roots are physical if
they are real. Moreover, if $m<0.96$, both physical roots are negative, 
$s<(u_{0}-\sqrt{u_{0}^{2}+4})/2$, and if $m>0.96$, both of them are positive, 
$s>(u_{0}+\sqrt{u_{0}^{2}+4})/2$. Because $s$ is the velocity of the
traveling CnWs, in these two cases we will call the waves left-moving and
right-moving ones, respectively.

Equation (\ref{dsdt}) also provides for a necessary, although, strictly
speaking, not sufficient, information about the stability of the CnW
solutions. Indeed, a necessary condition is that the corresponding FP be
stable within the framework of eq. (\ref{dsdt}). It can be checked
that the sign of r.h.s. of eq. (\ref{dsdt}) always changes from positive to
negative when passing (from left to right) through FP corresponding to a larger
value of the speed $\left| s\right| $ (it can be demonstrated that it always
corresponds to a larger value of the amplitude $\eta $ too), so this FP is
obviously stable as a solution to eq. (\ref{dsdt}). Similarly, FP
corresponding to the smaller values of $\left| s\right| $ and $\eta $ is
always unstable, playing the role of a separatrix bordering the attraction
domain of the stable FP. Direct simulations (see below) have demonstrated
that, whenever the CnW solution corresponds to FP is stable as the solution
to eq. (\ref{dsdt}), it turns out to be as well stable as a periodic-wave
solution of the full system of eqs. (\ref{u}) and (\ref{v}).

It is relevant to mention that quite a similar conclusion was conjectured
analytically and proved by means of direct simulations in ref. \cite
{Malomed01} for SW solutions to eqs. (\ref{u}) and (\ref{v}). The SW
solutions also appeared in pairs, the pulses with larger and smaller values
of the amplitude being stable and unstable, respectively.

Note that all the above conditions, for instance, eqs. (\ref{cond1}),
involve only the mean value $u_{0}$ of the field in the main ($u$-)
subsystem, but not the mean value $v_{0}$ of the extra field. Therefore, we
conjecture that the stability of the CnW solutions 
does not depend on $v_{0}$. 
This point will be confirmed by numerical simulations reported in the next
section. Thus, for fixed values of the gain and loss parameters $\alpha $, 
$\gamma $ and $\Gamma $, the stability of the periodic waves is determined by 
$u_{0}$ and $m$, the latter constant being related to the period $L$, see
eqs. (\ref{KdV-etaL}). This suggests to display stability regions for the
CnW solutions in the ($L$, $u_{0}$) parametric plane, if the values of 
$\alpha $, $\gamma $ and $\Gamma $ are fixed, which will be done below.

Adopting the principle stated above, i.e., that the CnW solution with the
larger values of $\left| s\right| $ and $\eta $ is stable, we can summarize
predictions for the stability in the form of parametric regions; in the next
section, the predictions will be checked against direct simulations of eqs. 
(\ref{u}) and (\ref{v}). Figure 1 shows the stability region predicted in the
plane ($L,u_{0}$) for the case $\alpha =0.15$, $\gamma =0.05$, and $\Gamma
=0.2$. Three different areas can be identified in this case: the
right-moving waves are stable in the area I, constant solutions are stable
in the area II, and the left-moving waves are stable in the area III. It can
be seen from Fig. 1 that, within the range considered, the left-moving waves
disappear if the mean value $u_{0}$ exceeds a certain critical value 
($\approx 0.6$), whereas the right-moving waves only exist when $u_{0}$ is 
{\em larger} than another critical value ($\approx -0.12$).

Note that, when $u_{0}$ is fixed to be zero, Fig. 1 has gaps on the axis of 
$L$ inside which stable waves are absent. Indeed, the left-moving waves exist
if $4.0<L<6$ or $L\geq 8.0$, and the right-moving ones exist if 
$L\geq 16.75$, i.e., the gaps are $L<4.0$ and $6.0<L<8.0$. This situation can be
explained in detail in terms of eq.(\ref{cubic}). Indeed, it is easy to
check that, if $u_{0}=0$, two physical roots to eq. (\ref{cubic}) exist if 
$m\leq 0.7052$ or $m\geq 0.999858$, the nonexistence gap $6.0<L<8.0$ in Fig. 1 (for $u_{0}=0$) being a consequence of the discontinuity in the values of 
$m$ at which eq. (\ref{cubic}) has two physical roots. More accurately, 
$m=0.7052$ corresponds to $L=5.997$. It is the maximum value of $L$ for which
the left-moving periodic waves with a {\em single maximum} per period can be
found; such waves may be naturally called single-hump ones. Left-traveling
waves reappear at $L\geq 8.0$, this time having $N=2,3,...$ identical humps
inside one period (so that the fundamental period of the wave is $L/N$
rather than $N$). Single-hump right-moving waves appear when $L\geq 16.75$,
which corresponds to the above-mentioned value of $m=0.999858$. At this
point, a pair of complex conjugate roots of eq. (\ref{cubic}) bifurcate into
two physical ones.

On the other hand, eqs. (\ref{cubic}) and (\ref{a}) predict that, as 
$m\rightarrow 0$, the period $L$ takes the minimum value, $L_{\min }=$ 
$3.965$, at which stable CnWs may exist. 
This explains the presence of the other above-mentioned gap, $L<4$.

The model easily gives rise to a multistability, i.e., coexistence of
several attractors at the same values of parameters, an example being
overlapping between the areas I and III in Fig. 1. The multistability is
well confirmed by numerical simulations reported in the next section.

Figure 2 displays stability areas for another set of parameter values: 
$\alpha =0.1$, $\gamma =0.05$, $\Gamma =0.15$. In this case, there is no
overlapping between the areas I and III, which are separated by the area II,
where a constant solution is the single attractor. Moreover, no stable left-
or right-moving waves can be found for $u_{0}=0$ in this case, which is a
drastic difference from the situation displayed in Fig. 1. Left-moving waves
exist if $u_{0}$ is smaller than $-0.15$, while the right-moving waves cease
to exist if $u_{0}\leq 0.2$ and $L\leq 35$. 
In particular, when $u_{0}=-0.2$, the single-hump stable periodic 
wave exists in an interval $5.9<L<7.2$,
and when $u_{0}=-0.3$, the interval is $5.5<L<8.5$. For these values of 
$u_{0}$, multistability takes place at large values of $L$, not as the
coexistence of left-and right-traveling waves, but rather as coexistence of
stable waves traveling in one direction but having different numbers of
humps per period. As it was shown in ref. \cite{Malomed01}, the model also
supports a stable SW at the same values of the parameters. Therefore, a
conclusion is that, as $L\rightarrow \infty $, the stable single-hump
right-moving wave carries over into the stable SW,  with the corresponding mean
value $u_{0}$ vanishing in this limit.\\

\hspace*{-0.7cm}{\bf \S 4. Numerical Simulations}\\

In this section, we aim to verify predictions for the existence and
stability of periodic waves by direct simulations. To this end, eqs. 
(\ref{u}) and (\ref{v}) were integrated 
by an implicit Fourier pseudo-spectral
method (the same as the one used in ref. \cite{Malomed01}), with two
different types of initial conditions, $\sin (kx)$ and 
${\rm sech}^{2}(\kappa x)$.

Firstly, for the same set of parameter values, $\alpha =0.15$, 
$\gamma =0.05$, $\Gamma =0.2$, as was used in 
Fig. 1, and $u_{0}=0$, various numerical
experiments were conducted with different values of $L$ and different
initial conditions. As a result, stable periodic left-moving waves with a
single hump per period were found in an interval $4.0<L<6.5$, at it was also
concluded that stable waves are absent in an interval $6.5<L<7.9$.
Left-moving periodic waves with two identical humps per period were found
for $L\geq 7.9$. These findings are in very good agreement with the
predictions of the perturbation theory which were explained in detail in the
previous section and summarized in Fig. 1. On the other hand, single-hump
right-moving periodic waves were observed for $L\geq 18.0$, which should be
compared to the above-mentioned theoretical prediction $L>16.75$.

Figure 3 shows profiles of the stable waves found from the simulations for 
$L=6.0$ and $L=10.0$, with one and two humps per period, respectively. These
profiles were obtained from the $\sin (kx)$ initial conditions. At these
values of the parameters, all the initial profiles either evolve into a
single wave, or decay into a trivial state. It can be checked that the
numerically found values of the amplitude and velocity of the wave agree
well with the analytical predictions, see caption to Fig. 3.

It is also necessary to check the prediction that the mean value $v_{0}$ of
the extra field does not affect the stability of the periodic waves. To this
end, we conducted a numerical experiment, fixing $L=12$ and $u_{0}$ to be 
$2.0$ or $-1.0$. These values correspond to points in the areas I and III,
respectively, in Fig. 1. Then, we investigated the established wave profiles
for various values of $v_{0}$. A typical result is displayed in Fig. 4,
clearly showing that the mean value $v_{0}$ affects neither the stability
nor the shape of the $u$-component.

Another important issue to be checked is the multistability. Based on the
previous results, it is clear that the separation between local maxima in a
stable periodic left-moving wavetrain (for the same values of parameters as
in Figs. 3 and 4) is between $4.0$ and $6.0$, as predicted analytically, or
between $4.0$ and $6.5$ as found numerically, if $u_{0}=0$. Therefore, when
the period is $L=18.0$, the number $N$ of humps per period must belong to
the interval $18/6.5<N<18/4.0$, which yields $N=3$ or $N=4$. Thus, the
coexistence of two left-traveling waves with the fundamental periods $L/3$
and $L/4$ is expected in this case. As it was mentioned above, the stability
of the periodic waves with the fundamental period essentially smaller than 
$L $ implies that waves in the present model may also be stable against
perturbations breaking their periodicity. In fact, taking larger values of 
$L $, we were able to observe, at various values of the parameters, {\em
stable} periodic waves with the fundamental period as small as $L/16$.

The single-hump right-moving stable wave was predicted to exist too at the
same values of the parameters for $L\geq 18.0$. All these predictions were
confirmed by numerical simulations. Stable left-moving waves with three and
four humps per period, and stable single-hump right-moving wave are shown in
Fig. 5 for $L=18.0$. Again, the numerical results for the amplitudes and
velocities of the waves are found to agree well with the theoretical
predictions. We stress that no other waves but the ones mentioned here had
been turned up by simulations at these values of the parameters.

Lastly, we display numerical results for another set of parameters: $\alpha
=0.1$, $\gamma =0.05$, and $\Gamma =0.15$. When $u_{0}=0$, no left- or
right-traveling stable periodic wave can be found, various initial
conditions relaxing into a trivial uniform state in this case. Recall that
precisely this feature was predicted in the previous section (see also Fig. 2). 
With $u_{0}=-0.2$, stable single-hump left-moving waves are found for 
$5.8<L<7.2$ (the analytical prediction being $5.9<L<7.2$); when $u_{0}=-0.3$,
stable single-hump left-moving waves are found for $5.5<L<8.4$ (the
analytical prediction is $5.5<L<8.5$). As is seen, the numerical results
agree with the predictions very well. In other words, the inter-pulse
separation for a stable periodic left-traveling wavetrain is, in this case,
between $5.8$ and $7.2$ for $u_{0}=-0.2$, and between $5.5$ and $8.4$ for 
$u_{0}=-0.3$. Therefore, if the period $L$ is large, the multistability must
manifest itself as the coexistence of several different waves with multiple
humps. Figure 6 shows two- and three-hump left-moving waves for fixed values
of $L=24.0$ and $u_{0}=-0.3$. In this case, the possible number of the humps
per period may be $24/8.4<N<24/5.5$, implying $N=3$ or $N=4$, which is
confirmed by Fig. 6. On the other hand, when $u_{0}$ becomes positive, there
may exist right-moving single-hump waves. Several right-moving waves are
shown in Fig. 7 for different values of $u_{0}$ and $L$.\\

\hspace*{-0.7cm}{\bf \S 5.  The Zero-Dispersion Limit}\\

All the results presented above were obtained for the case when the linear
gain and loss terms in the underlying equations (\ref{u}) and (\ref{v}) are,
as a matter of fact, small perturbations in comparison with the dispersive
term (third derivative) in eq. (\ref{u}). It is also interesting to consider
the opposite limit of the zero-dispersion system, when the KS-KdV equation 
(\ref{u}) is replaced by the KS equation proper, 
so that the system takes the form 
\begin{eqnarray}
u_{t}+uu_{x}-v_{x} &=&-\alpha u_{xx}-\gamma u_{xxxx},  \label{KSu} \\
v_{t}-u_{x} &=&\Gamma v_{xx}, \label{KSv}
\end{eqnarray}
[following the arguments presented above, we set $c=0$ in eq. (\ref{KSv})].

First of all, it is relevant to mention that, on the contrary to the model
based on the KS-KdV (Benney) equation, the zero-dispersion model does not
support stable SWs. Our numerical simulations demonstrate that, with the
decrease of the coefficient $D$ in front of the third derivative in the
KS-KdV equation [in eq. (\ref{u}) this coefficient was $1$], there is a
finite critical value $D_{{\rm cr}}$, such that no stable pulses can be
found for $D<D_{{\rm cr}}$. However, in the coupled system (\ref{KSu}) and (\ref
{KSv}), stable periodic waves continue to exist, both with one and several
humps, and, unlike the KS equation proper (see, refs. \cite{Chaos,S01} and
references therein), the present system is {\em not} prone to generate
dynamical chaos.

Typical examples of stable periodic-wave solutions to the system (\ref{KSu})
and (\ref{KSv}) with one and two humps are displayed in Fig. 8. Quite
similar results were obtained at many other values of the parameters, while
spatiotemporal chaos has never been observed, although the search for chaos
in the system's parameter space was not exhaustive.\\

\hspace*{-0.7cm}{\bf \S 6.  Conclusion}\\

In this work, we have studied cnoidal-like waves and their stability in the
system of linearly coupled KS-KdV equations. We first constructed a family
of cnoidal-wave solutions for the zero-order system, which does not include
the gain and loss terms. Then, treating the gain and dissipation as small
perturbations, and making use of the balance equation for the net field
momentum, we have found necessary conditions for the existence of stable
cnoidal waves. Finally, the predictions were checked against direct
numerical simulations, showing very good quantitative agreement in all the
cases considered. The obtained results can be summarized as follows:

\begin{enumerate}
\item  For mean value $v_{0}$ of the field in the linear subsystem does not
affect stability of the periodic waves. The stability is determined by the
mean value $u_{0}$ of the field in the main subsystem and the wave's period 
$L$.

\item  For fixed values of the model's parameters, three different stability
regions are found in the ($L$, $u_{0}$) plane, viz., region I for
right-moving waves, region II for trivial solutions and region III for
left-moving waves.

\item  The three stability regions can overlap with each other, which leads
to multistability.

\item  At large values of $L$, the multistability manifests itself also as
coexistence of stable waves moving in one direction but having different
fundamental periods $L/N$, with an integer $N$. In fact, $N$ may take large
values, which implies that the periodic waves are also stable against
perturbations breaking their periodicity.
\end{enumerate}

\newpage

\begin{center}
{\bf Figure Captions}
\end{center}

Fig. 1. Stability areas in the parametric plane ($L$,$u_{0}$) for $\alpha
=0.15$, $\gamma =0.05$ and $\Gamma =0.2$, as predicted by the perturbation
theory. Right-moving cnoidal waves are stable in the region I, which lies
above the solid curve, left-moving cnoidal waves are stable in the region
III, which is below the dashed curve, and the constant solutions are the
only stable solutions predicted in the region II.\\

Fig. 2. The same as in Fig. 2 for $\alpha =0.1$, $\gamma =0.05$ and $\Gamma
=0.15$.\\

Fig. 3. Numerically found profiles of stable periodic waves for $\alpha =0.15
$, $\gamma =0.05$, $\Gamma =0.2$ and $u_{0}=0$ (one period is displayed).
The solid and dashed curves show the $u$- and $v$-components: (a) $L=6.0$;
(b) $L=10.0$. In these two cases, numerically found amplitudes of the 
$u$~-~component are, respectively, $A_{u}=3.46$ and $A_{u}=5.85$, while the
perturbation theory predicts $A_{u}=3.65$ and $A_{u}=5.95$.\\

Fig. 4. Numerically found stable waves for $\alpha =0.15$, $\gamma =0.05$, 
$\Gamma =0.2$, $L=12.0$ and different values of $v_{0}$. The solid curve is
the profile of the $u$-component, while the dashed and dashed-dotted lines
are profiles of the $v$-component obtained for different $v_{0}$ (the $u$%
-component is shown by a single curve, as the change of $v_{0}$ produces no
visible variation in it): (a) $u_{0}=2.0$, with $v_{0}=0$ (dashed line) or 
$v_{0}=2.0$ (dot-dashed line); (b) $u_{0}=-1.0$, with $v_{0}=-1.0$ (dashed
line) or $v_{0}=0$ (dot-dashed line).\\

Fig. 5. Multistability of stable periodic waves for $\alpha =0.15$, $\gamma
=0.05$, $\Gamma =0.2$, $u_{0}=0,$ and $L=18.0$. The solid and dashed curves
represent the profiles of the $u$- and $v$-components. Panels (a) and (b)
show the left-moving waves with $3$ and $4$ maxima per period, and panel (c)
shows the right-moving wave with the single maximum.\\

Fig. 6. The same as in Figs. 5(a) and 5(b) for $\alpha =0.1$, 
$\gamma =0.05$, $\Gamma =0.15$ and $L=24.0$.\\

Fig. 7. Stable right-moving cnoidal waves for $\alpha =0.1$, $\gamma =0.05$
and $\Gamma =0.15$. The solid and dashed curves show the $u$- and 
$v$~-~components for (a) $L=12.0,\:u_0=2.0$, (b) $L=15.0,\:u_0=1.0$, 
(c) $L=30.0,\:u_0=0.4$. The numerically
found and analytically predicted values of the amplitude of the 
$u$~-~component in these three cases are, respectively, $A_{u}=5.48$ and $5.70$; 
$A_{u}=4.50$ and $4.68$; $A_{u}=5.23$ and $5.42$.\\

Fig. 8. Typical examples of stable periodic waves with one (a) and two (b)
humps per period, found by means of direct simulations of the
zero-dispersion model based on Eqs. (\ref{KSu}) and (\ref{KSv}), for $\alpha
=0.15$, $\gamma =0.05$, $\Gamma =0.2$, and $L=10$ (a) or $L=11$ (b). In both
cases, the mean value of the $u$-field is $u_{0}=-2$. The profiles of the 
$u$~-~field are shown.

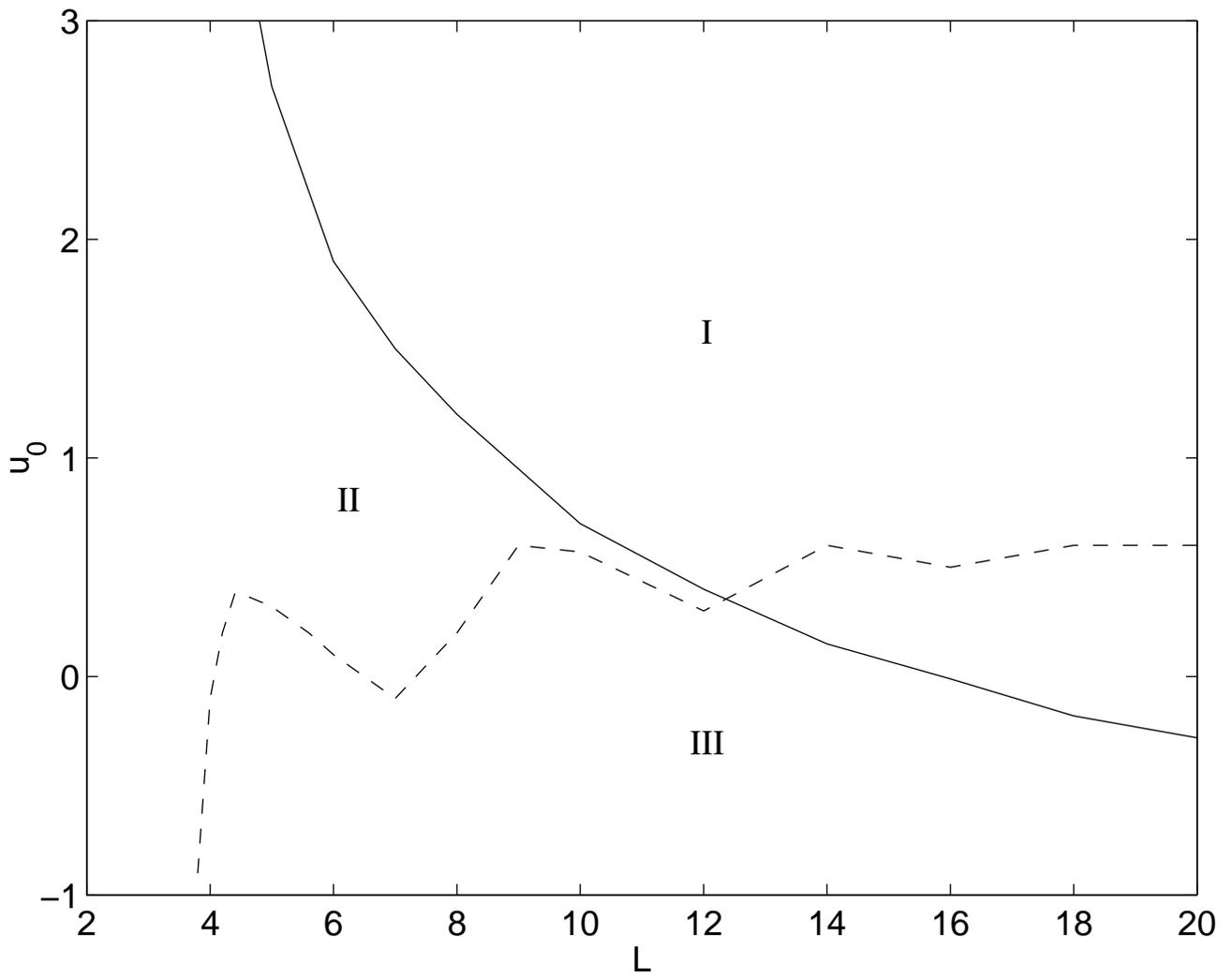
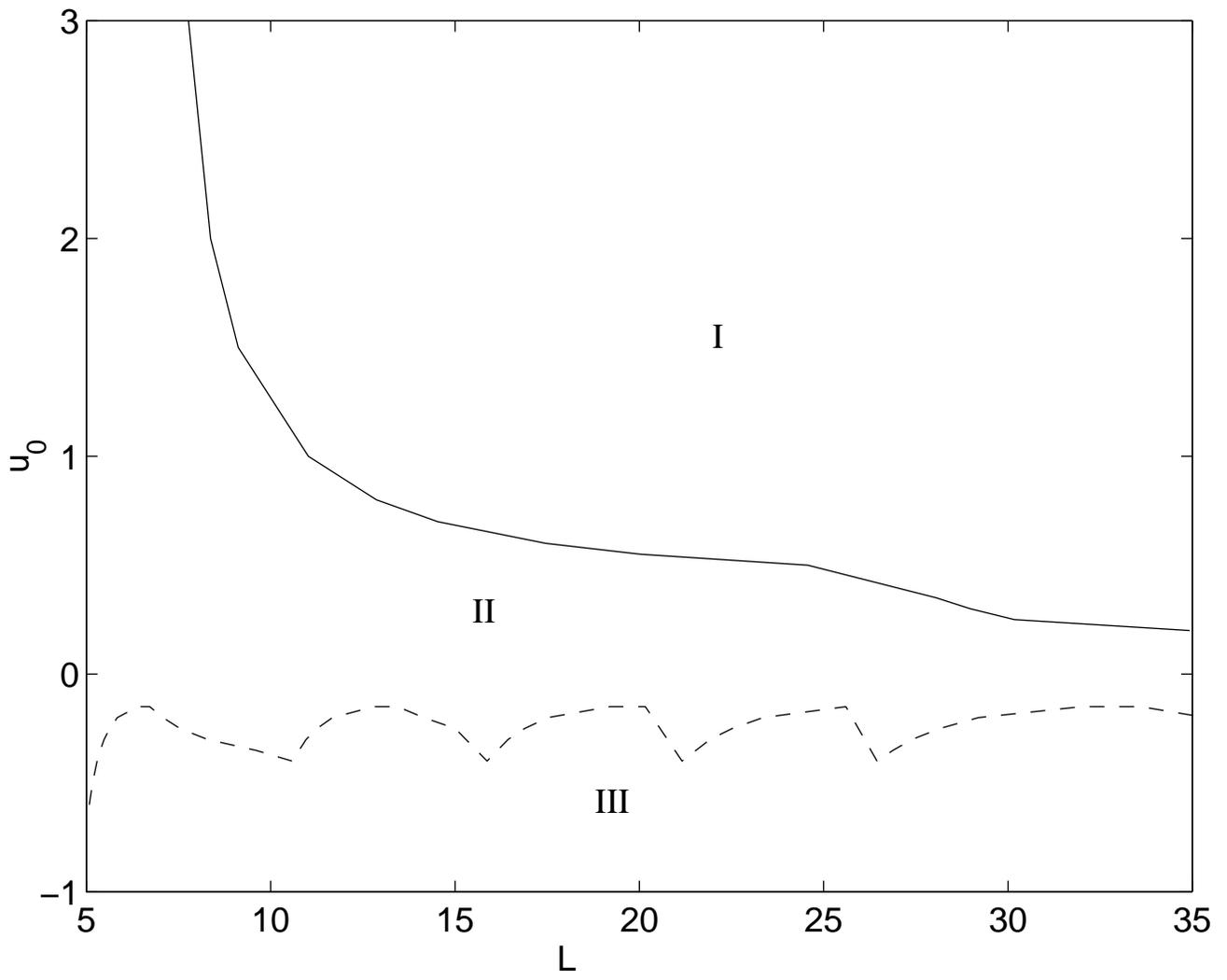
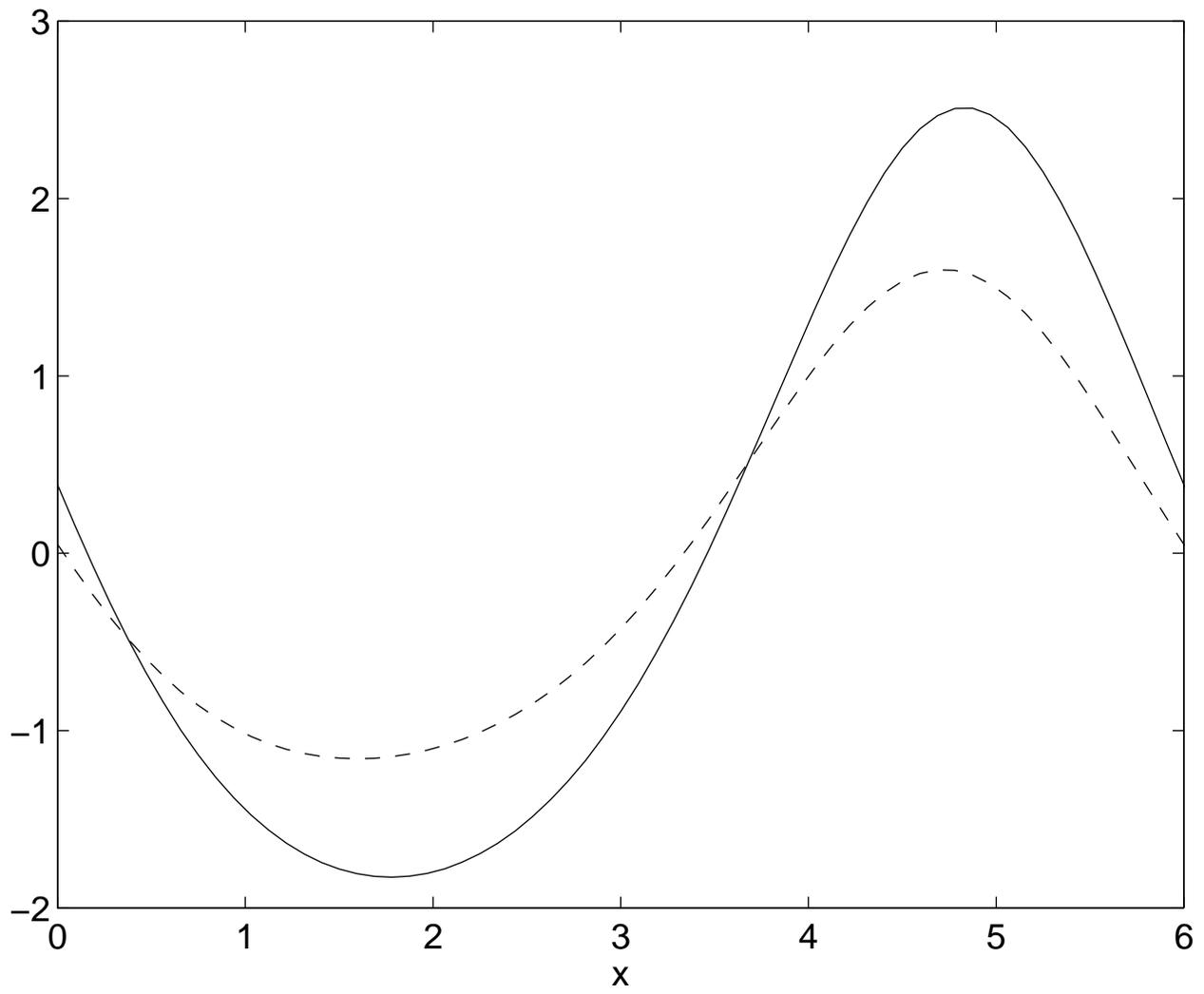
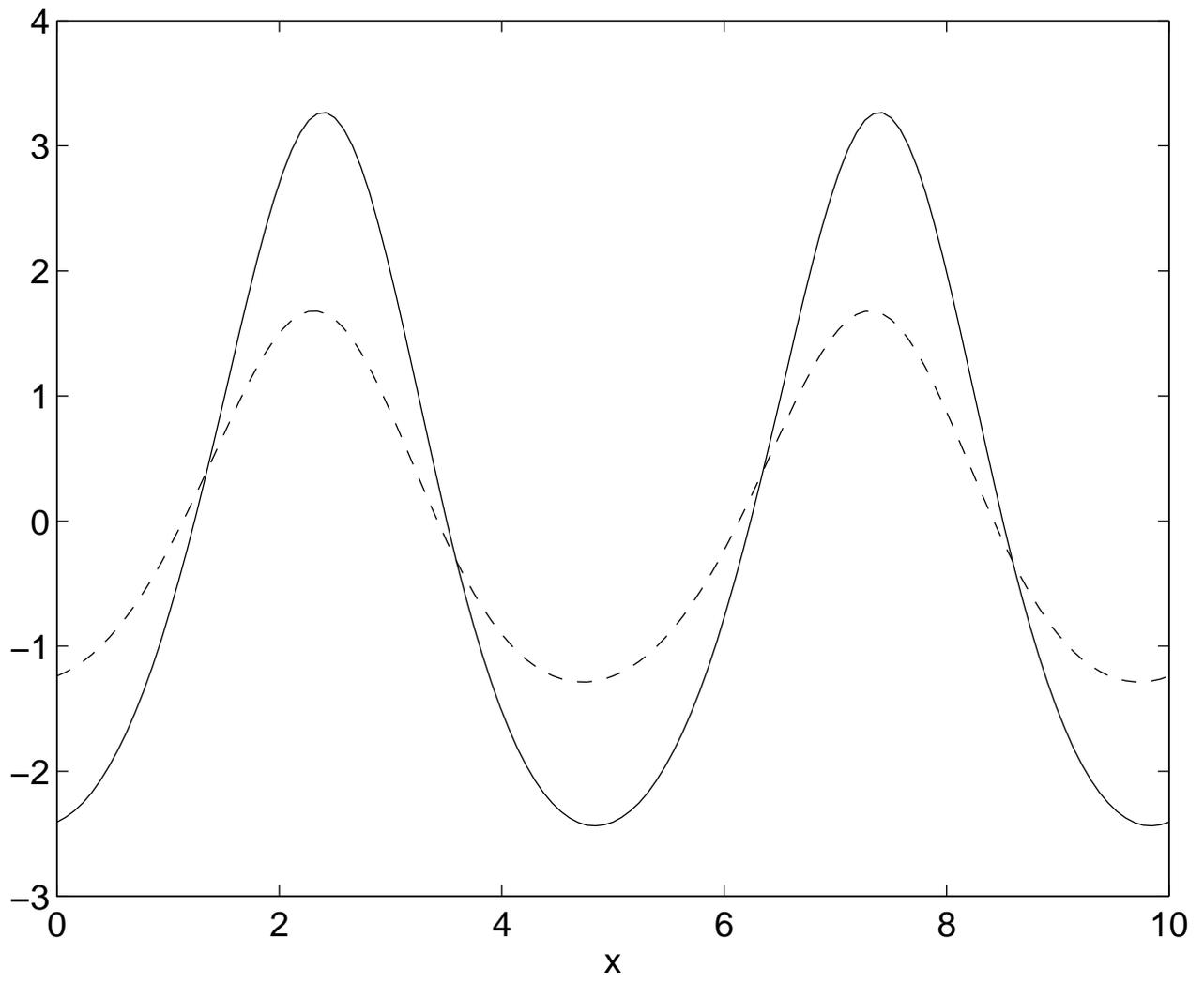
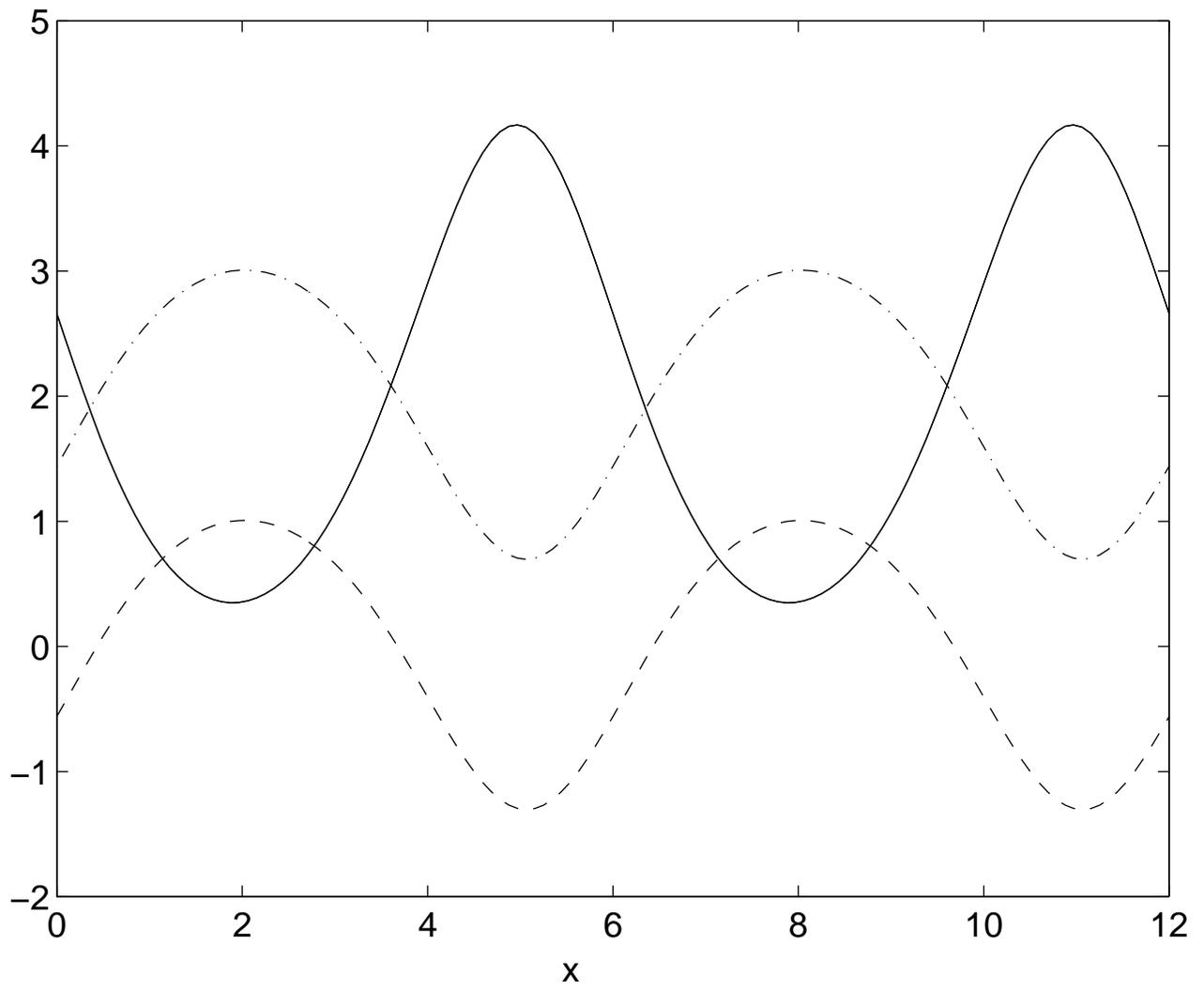
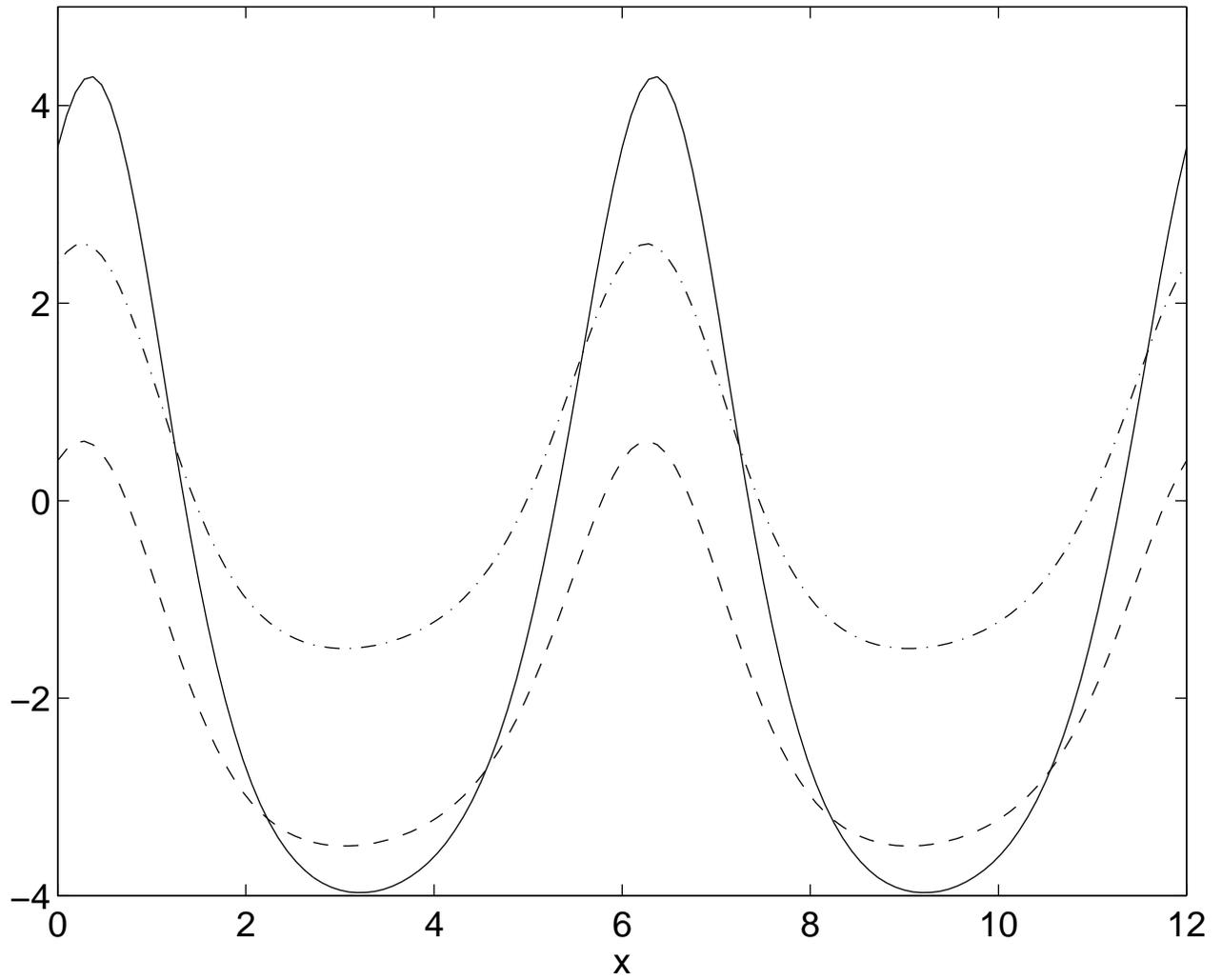
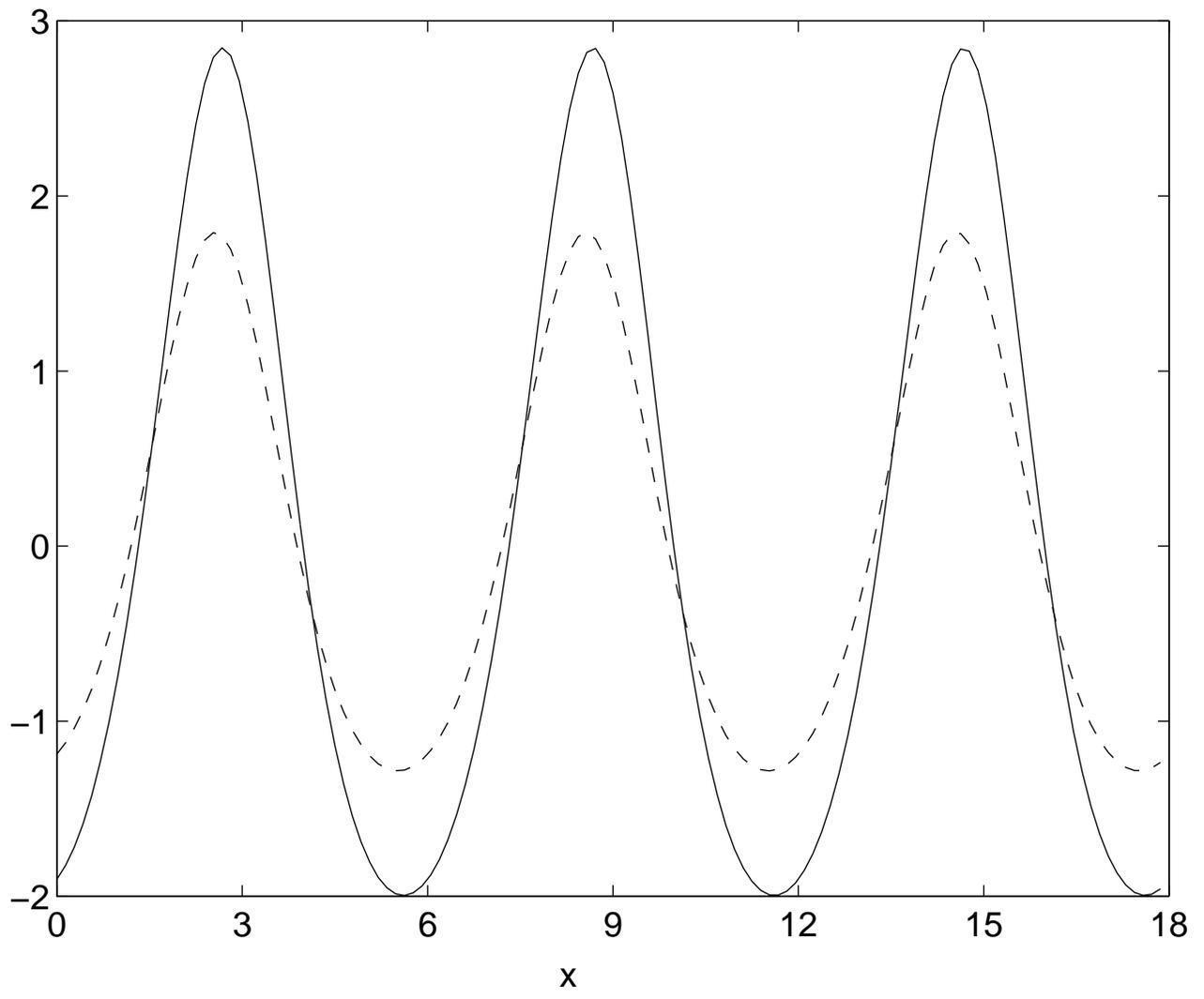
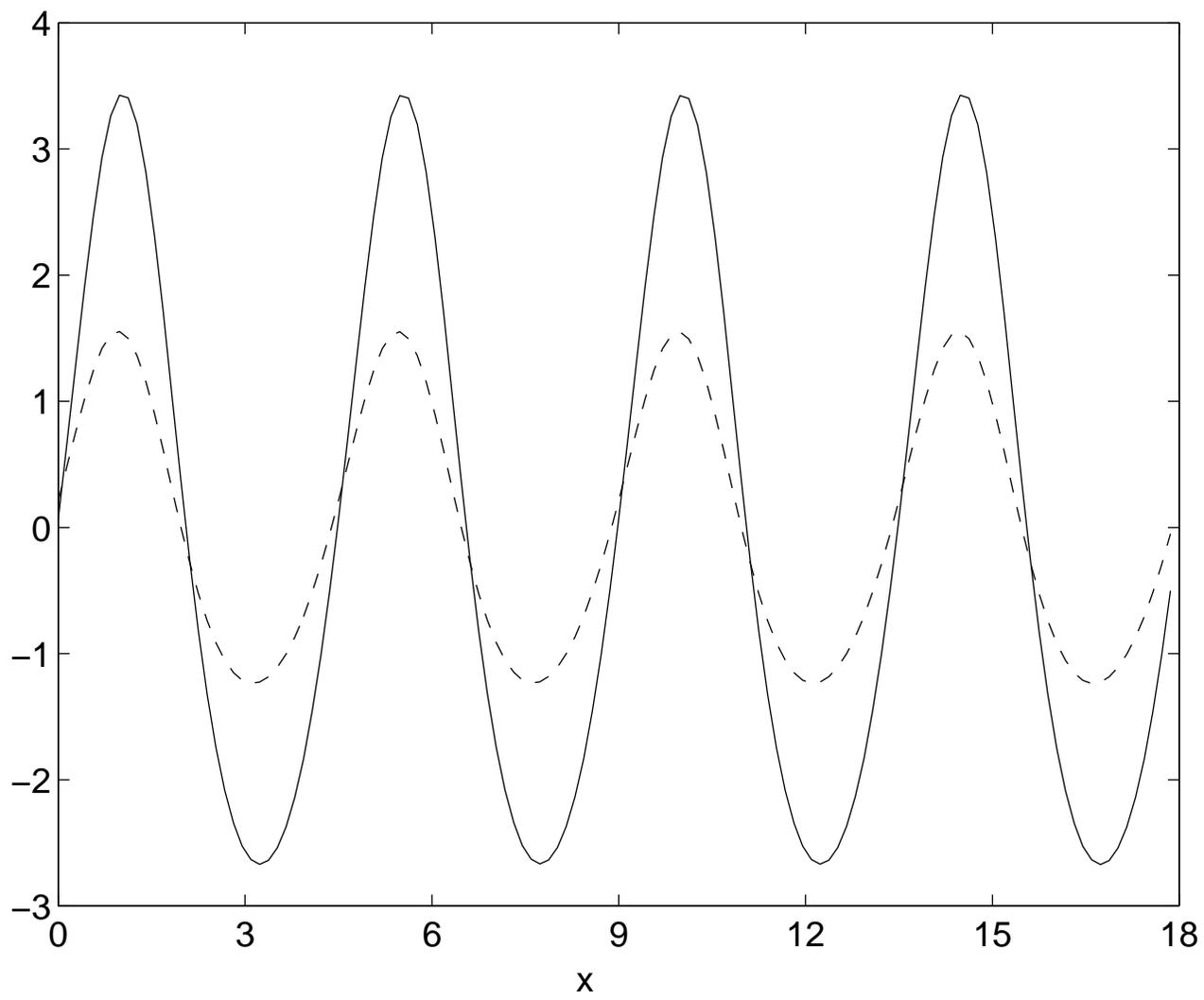
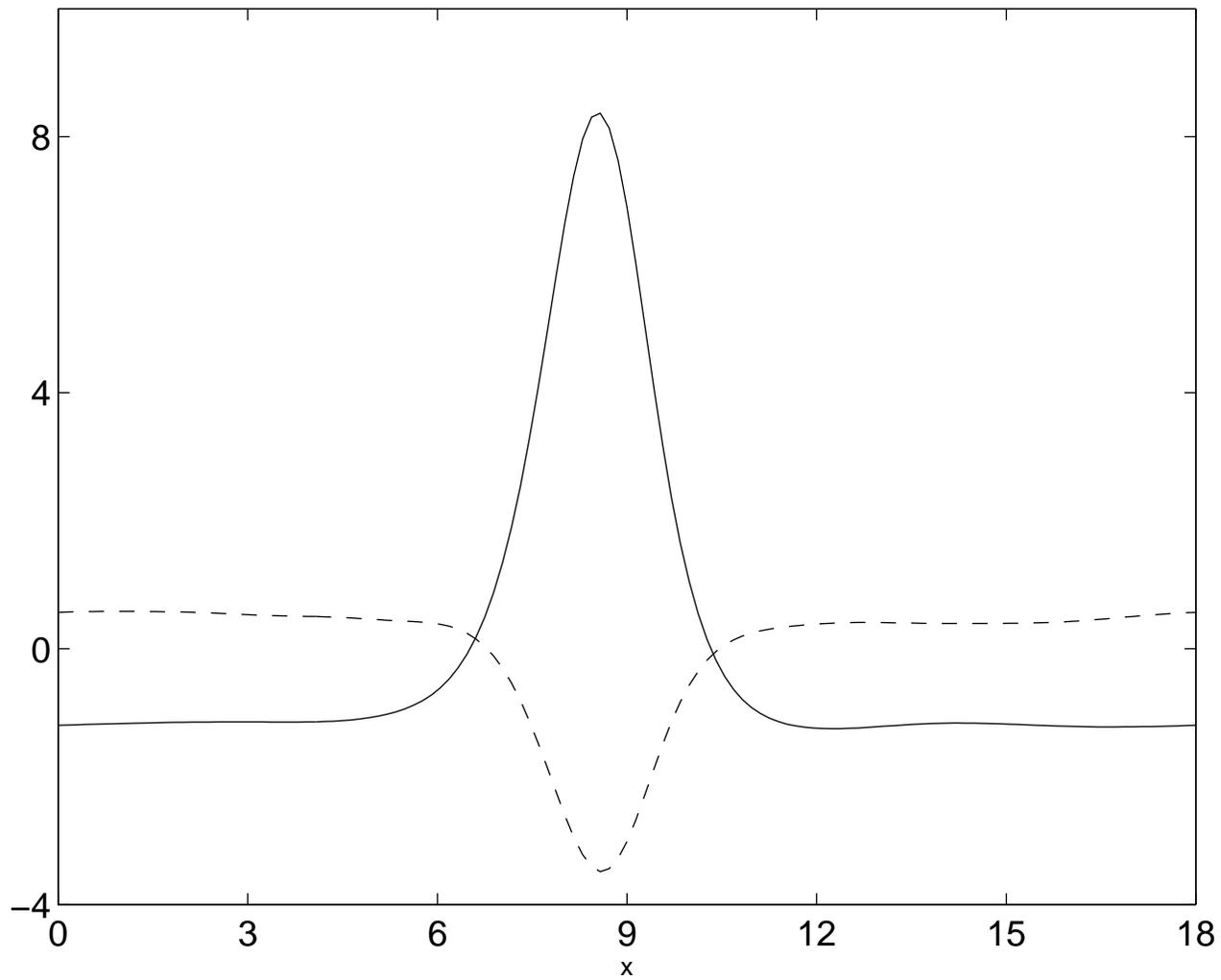
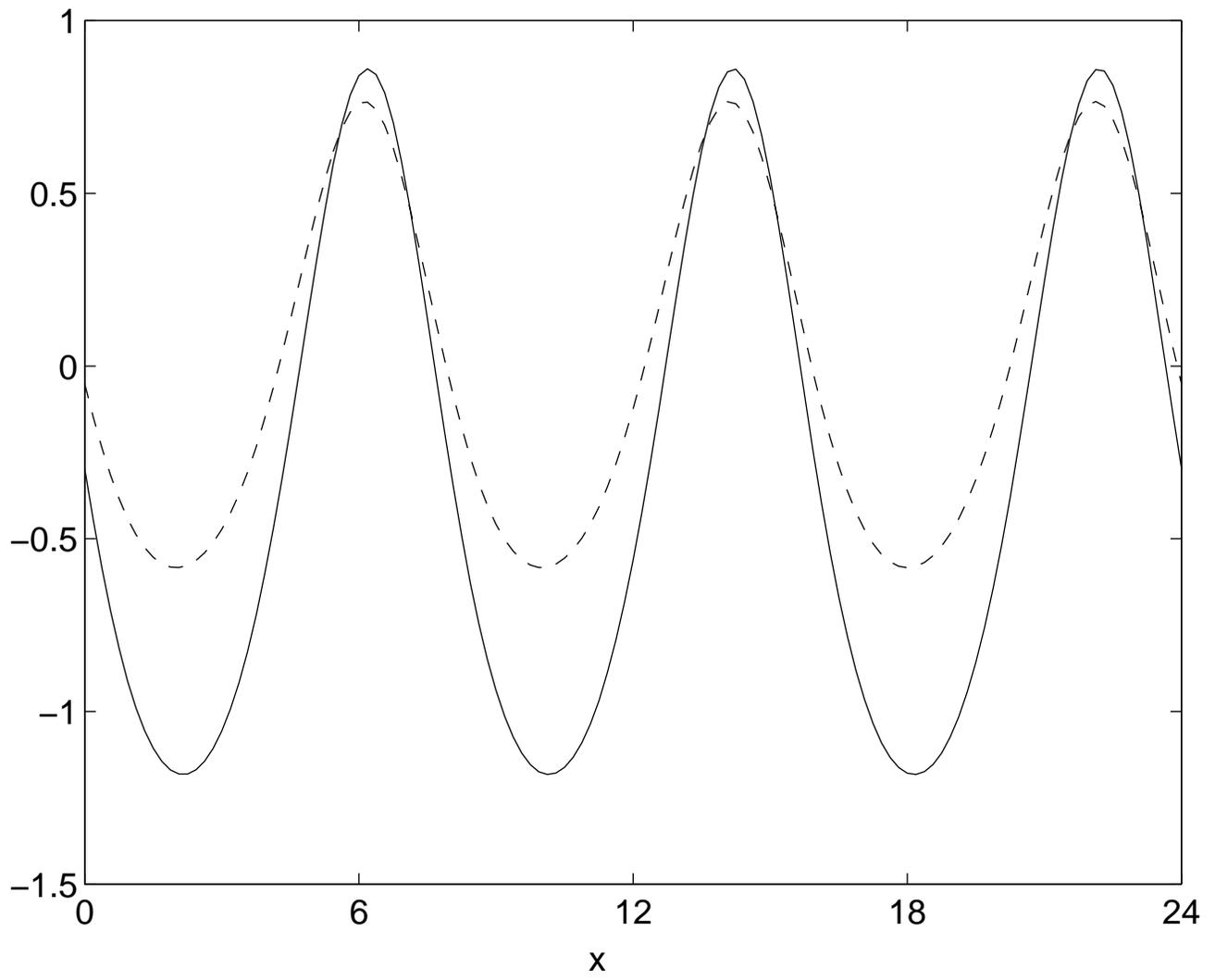
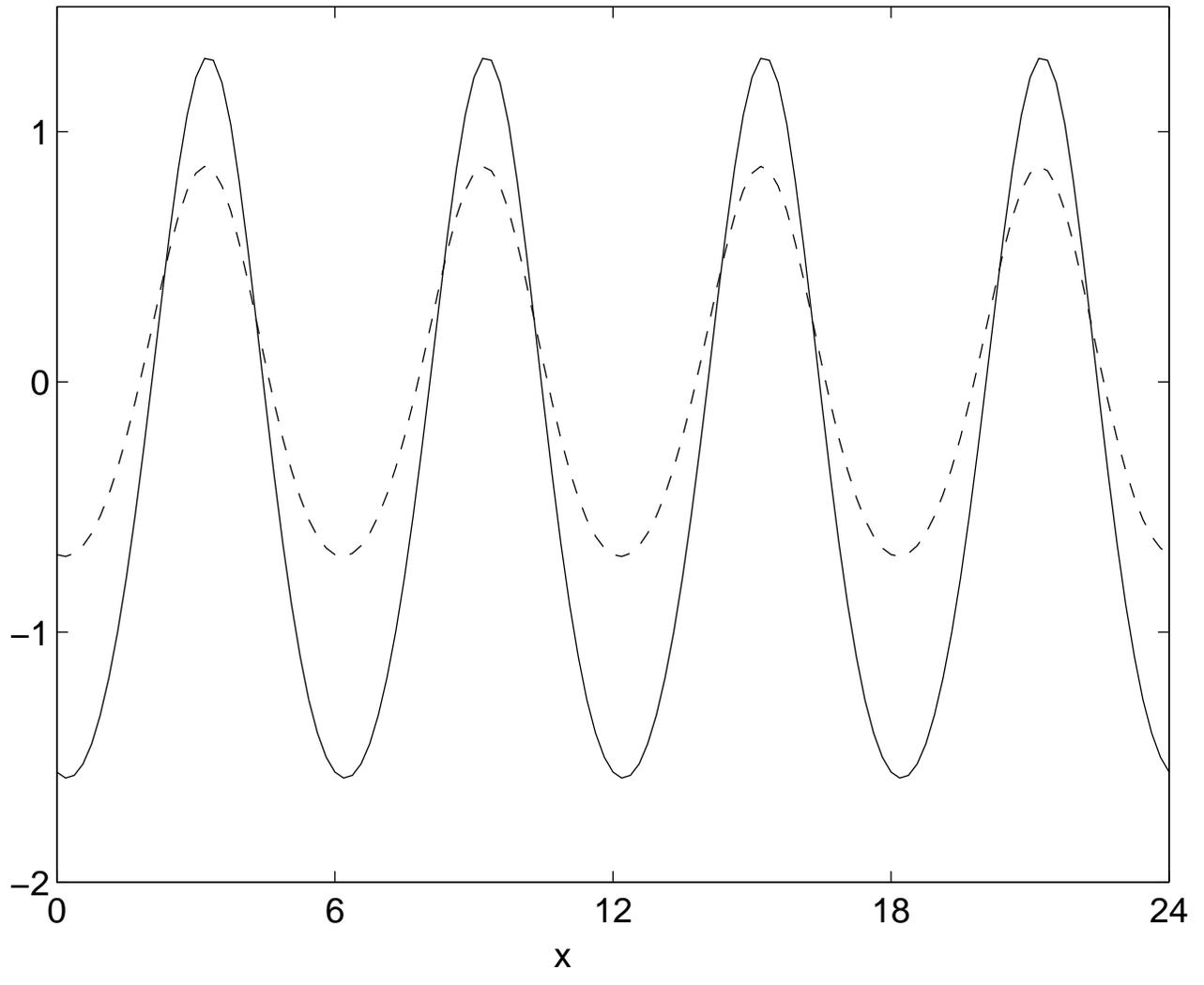
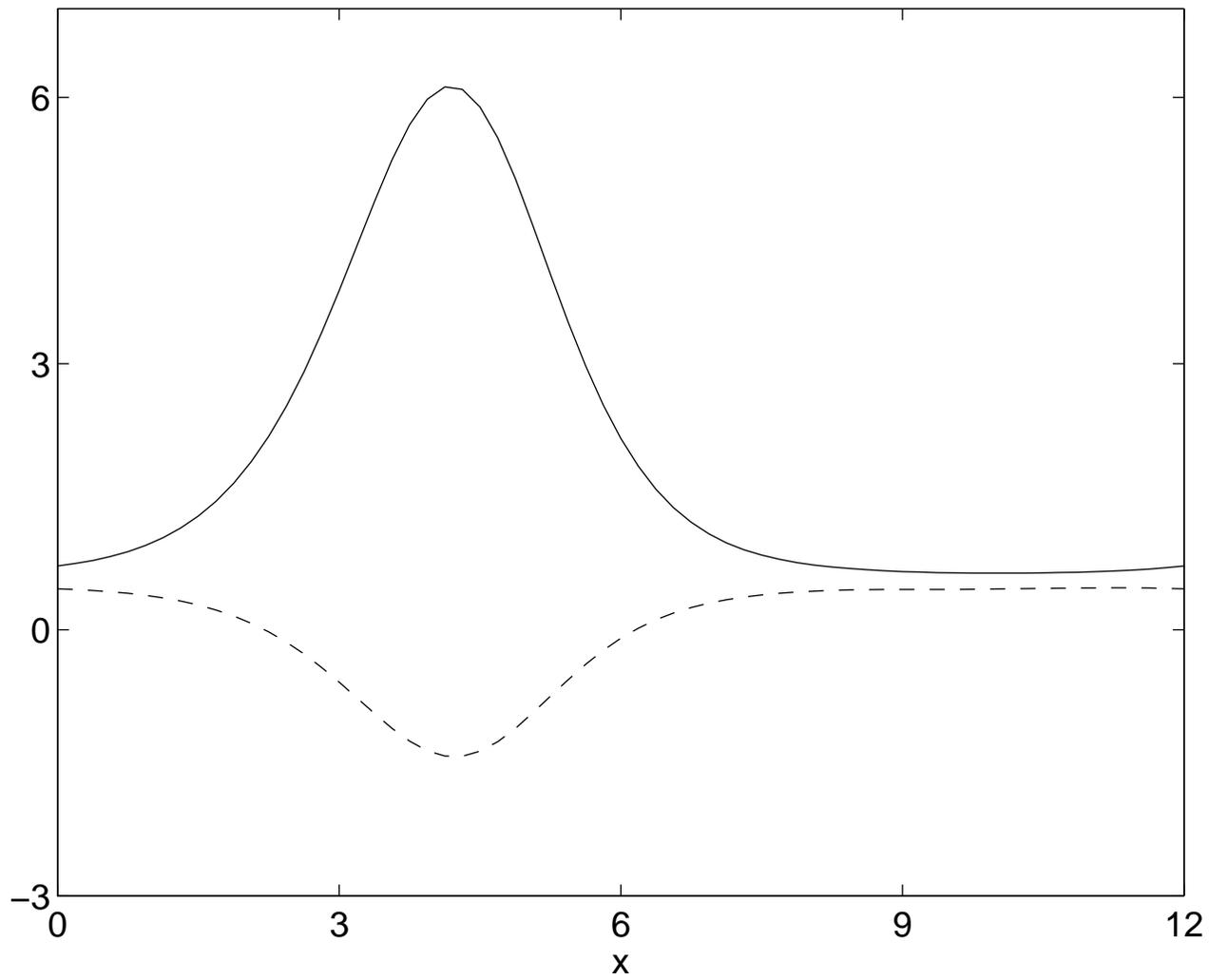
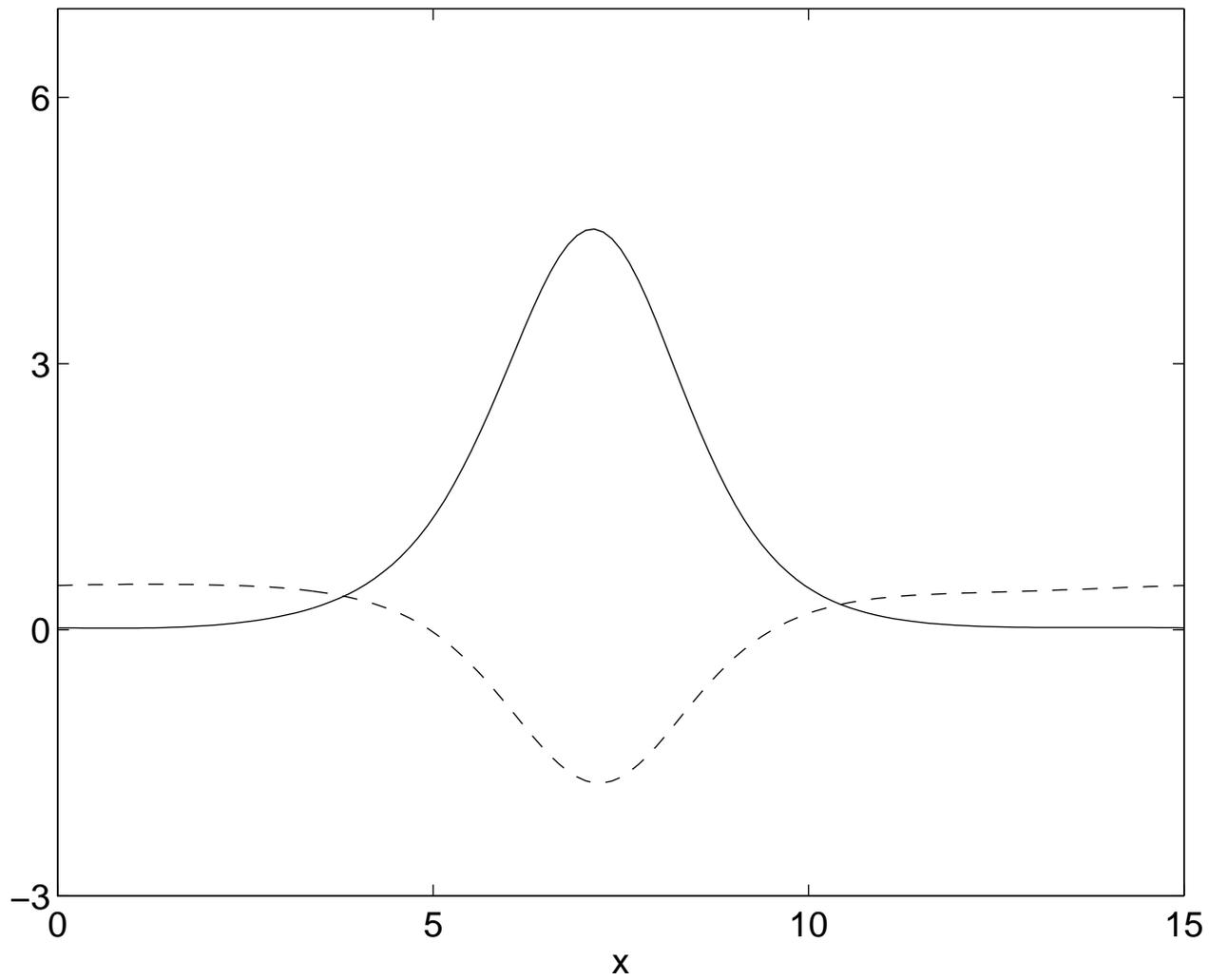
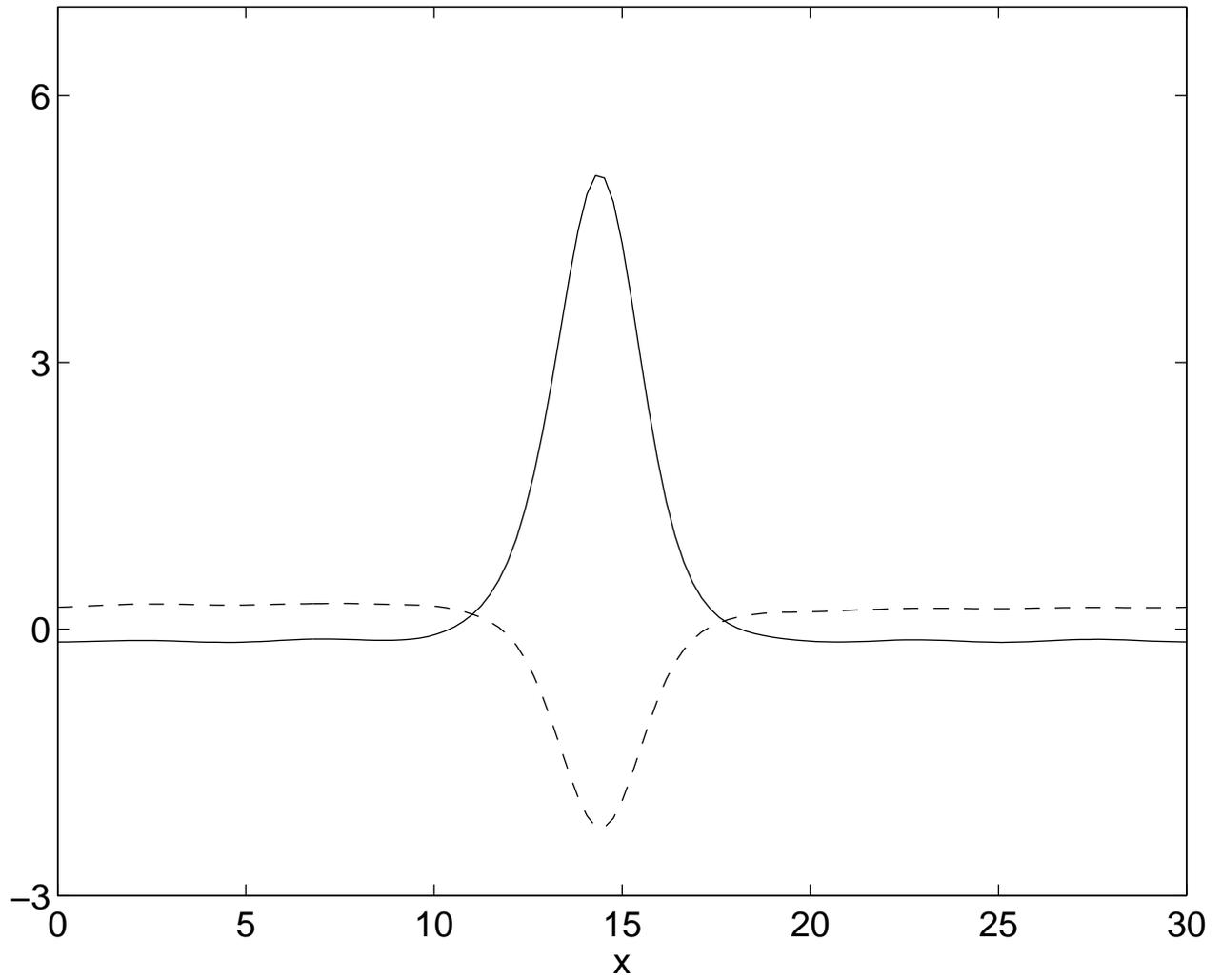
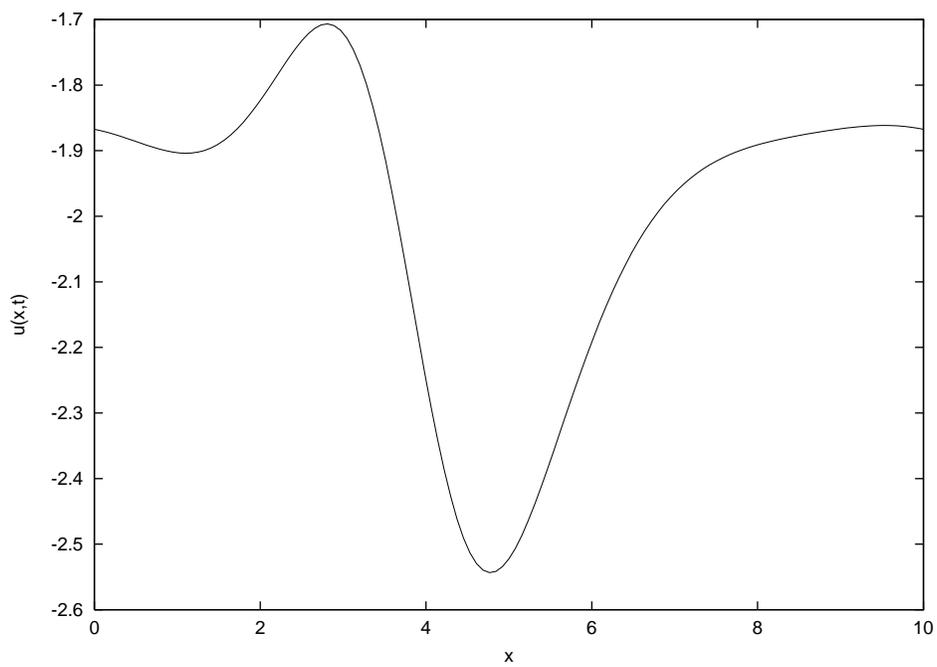
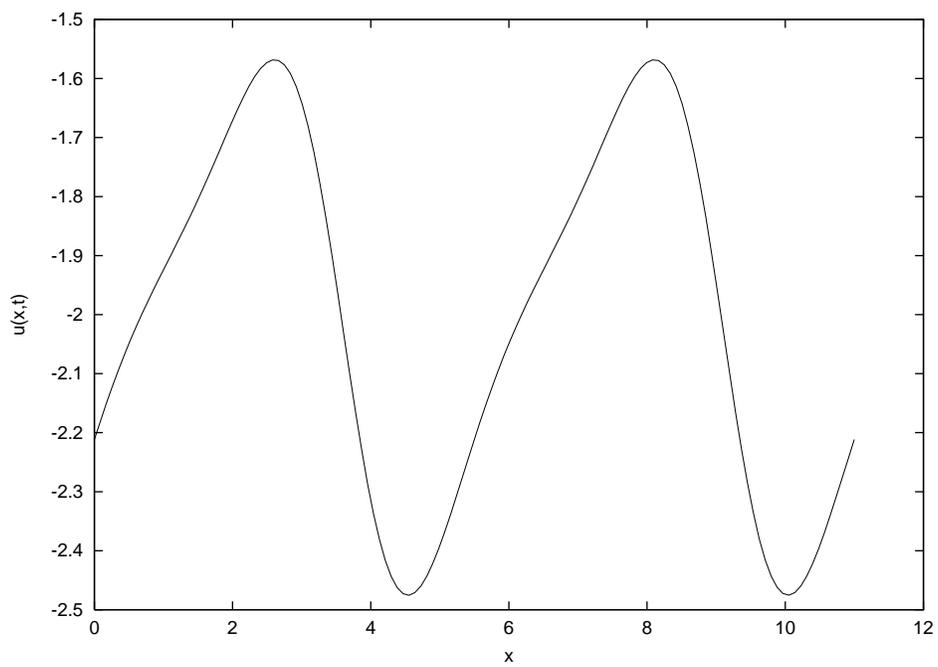
\end{document}